\newcommand{\al}{\mbox{$^{26}$\hspace{-0.2em}Al}}

\newcommand{\nuc}[2]{\mbox{$^{#1}$#2}}
\newcommand{\Msol}{\mbox{$M_{\sun}$}}

\newcommand{\gray}{\mbox{$\gamma$-ray}}
\newcommand{\pcmq}{\mbox{cm$^{-2}$}}

\newcommand{\psec}{\mbox{s$^{-1}$}}

\newcommand{\funit}{\mbox{ph \pcmq \psec}}

\def\MeV{\mbox{Me\hspace{-0.1em}V}}

\def\deg{\ensuremath{^\circ}}
\def\sun{\hbox{$\odot$}}

\documentclass[11pt]{article}
\usepackage{moriond,epsfig}

\def\Journal#1#2#3#4{{#1} {\bf #2}, #3 (#4)}

\begin{document}
\vspace*{4cm}
\title{SPI SCIENCE PROSPECTS}

\author{J. KN\"ODLSEDER and J.-P. ROQUES}

\address{Centre d'Etude Spatiale des Rayonnements, B.P. 4346, 31028 
Toulouse, FRANCE\\ 
E-mail: knodlseder@cesr.fr}

\maketitle

\abstracts{
With the launch of ESA's {\em INTEGRAL} satellite in october 2002,
a gamma-ray observatory will become available to the scientific community 
that combines imaging and spectroscopic capacities in the 
20 keV to 10 \MeV\ energy range.
In this paper, we summarise the science prospects of the SPI 
spectrometer aboard {\em INTEGRAL} which provides unprecedented 
spectral resolution with good imaging capabilities and high 
sensitivity.
Emphasise will be given to the key objectives of SPI which are 
the determination of the galactic positron origin,
the study of galactic star formation and nucleosynthesis activity via 
the radioactive trace isotopes \al\ and \nuc{60}{Fe},
and the study of nucleosynthesis in supernova explosions and the 
gamma-ray line emission from their remnants.
}

\section{The SPI telescope on INTEGRAL}
\label{sec:spi}

Due to continuing progress in instrumentation, the field of gamma-ray 
line astronomy has become a new complementary window to the universe.
With the COMPTEL and OSSE telescopes onboard {\em CGRO}, the sky 
has been imaged for the first time in the light of gamma-ray lines, 
leading to maps of 511 keV annihilation radiation and \nuc{26}{Al} 1.809 MeV 
emission.
New gamma-ray lines have been discovered, such as the 1.157 MeV line 
from \nuc{44}{Ti} or several decay lines from \nuc{56}{Co} and 
\nuc{57}{Co}.
Gamma-ray lines probe aspects of nucleosynthesis, stellar evolution, and 
supernova physics that are difficult to access by other means.
Additionally, they provide tracers of galactic activity and improve 
our understanding of the interstellar recycling processes.

With the upcoming {\em INTEGRAL} satellite, foreseen for launch in october 
2002, ESA provides a gamma-ray observatory to the scientific community that 
combines imaging and spectroscopic capacities in the 20 keV to 
10 \MeV\ energy range.
{\em INTEGRAL} is equipped with two gamma-ray telescopes, optimised for 
high-resolution imaging (IBIS) and high-resolution spectroscopy 
(SPI), supplemented by two X-ray monitors (JEM-X) and an optical 
monitor (OMC).
With respect to precedent instruments, the {\em INTEGRAL} telescopes 
provide enhanced sensitivity together with improved angular and 
spectral resolution.
In particular, SPI will map gamma-ray lines with an angular resolution of 
about $2\deg$ and a spectral resolution of $E / \Delta E \sim 500$ 
(at 1 \MeV), corresponding to Doppler velocities of $\sim 600$ km s$^{-1}$.
Consequently, gamma-ray line astrophysics figures among the prime objectives 
of this telescope.

SPI consists of a pixelised gamma camera with a geometrical area 
of $\sim500$ cm$^{-2}$, made of 19 hexagonal high-purity germanium 
crystals cooled actively to a temperature of $\sim85$ K.
Imaging capabilities are achieved by placing a coded mask at about
$1.7$ meters above the detector plane, providing the spatial 
modulation of the incoming gamma-rays that allow the reconstruction of 
the underlying source intensity distributions.
The mask is made of 3 cm thick tungsten elements which provide the 
necessary absorption efficiency up to high gamma-ray energies.
The wide field of view of $16\deg$ (fully coded; 34\deg\ partially 
coded) is defined by an active shield made of BGO scintillator 
crystals which also acts as anticoincidence shield for instrumental 
background reduction.
A plastic scintillator placed under the mask aims in reducing the 
511 keV instrumental background line, improving the telescope's 
sensitivity at this astrophysically important energy.
A pulse shape discrimination electronics provides further background 
reduction in the 200 keV - 2 \MeV\ domain, which results in an 
expected (narrow) gamma-ray line sensitivity of
$5 \times 10^{-6}$ \funit\ ($3\sigma$) at 1 \MeV\ for an observing time 
of $10^6$ seconds.

In the following sections, key science topics that can be addressed 
with these performances will be presented.
Emphasise will be given to gamma-ray line spectroscopy since the SPI 
telescope has been optimised for such studies.
However, with its excellent continuum emission sensitivity, in 
particular in combination with the large field of view and moderate 
angular resolution, SPI is also well suited for studying continuum 
sources, in particular if they are of diffuse nature.

\section{Stellar nucleosynthesis}
\label{sec:massive}

\subsection{Key questions}

Stars more massive than $\sim8$ \Msol\ are the most prolific 
nucleosynthesis sites in the universe.
During their short lives they synthesise large quantities of heavy 
elements that enrich the interstellar medium either through stellar wind 
ejection or at the final explosion of the star in a supernova event.
Although the big picture of element synthesis is already understood 
since the 50ies \cite{burbidge57}, many details are still poorly known, 
and theoretical yield predictions generally suffer from large 
uncertainties \cite{woosley99}.
In particular, the physics of mixing processes within massive stars 
is not well understood, and the impact of stellar rotation and/or 
close binary evolution can substantially alter abundance 
patterns \cite{meynet00}.
By measuring isotopic nucleosynthetic yields using gamma-ray line 
observations, SPI can provide important constraints on the mixing 
processes, and may provide clues on the effects of stellar 
rotation \cite{knoedl02}.
Also, mass loss through stellar winds is crucial for the evolution 
of the most massive stars, and gamma-ray line observations of mass-losing 
stars (in particular Wolf-Rayet stars) allow a direct study of its impact 
on nucleosynthesis yields \cite{oberlack00}.

Similar to the use of radioactive tracers in medicine, freshly 
produced radioactive isotopes trace also the distribution of 
nucleosynthesis activity throughout the Galaxy, and gamma-ray line 
observations may be employed to study this activity.
As we will show in the next section, the radioactive isotope \al\ is 
an excellent candidate tracer for this purpose, complementing other 
tracers of massive star activity such as the molecular gas distribution, 
far-infrared emission or free-free emission from the ionised interstellar 
medium.

\subsection{Galactic structure and distribution}

With the detection and the mapping of the 1.809 \MeV\ gamma-ray line 
from \al, gamma-ray line astronomy has made important progress during 
the last 15 years.
The COMPTEL telescope aboard {\em CGRO} has 
provided the first all-sky map of this radioactive isotope, which with 
a lifetime of $\sim10^6$ years is an unambiguous proof that nucleosynthesis 
is still active in our Galaxy \cite{oberlack96}.
The 1.809 \MeV\ map shows the galactic disk as the most prominent emission 
feature, demonstrating that \al\ production is clearly a galaxywide 
phenomenon.
The observed intensity profile along the galactic plane 
reveals asymmetries and localised emission enhancements, 
characteristic for a massive star population that follows the 
galactic spiral structure.
Thus, \al\ production seems related to the massive star population.
Correlation studies using tracer maps for various source candidate 
populations strongly support this suggestion, which in particular 
revealed that 1.809 \MeV\ gamma-ray line emission follows closely the 
distribution of galactic free-free emission which is powered by the 
ionising radiation of stars with initial masses $>20$ \Msol\ 
\cite{knoedl99}.
This suggests that explosive \al\ production in supernovae may be less 
important than previously thought and hydrostatic nucleosynthesis in massive 
mass-losing stars may possibly be the primary production channel for galactic 
\al.

\begin{figure*}
   \epsfxsize=7.5cm \epsfclipon
   \epsfbox{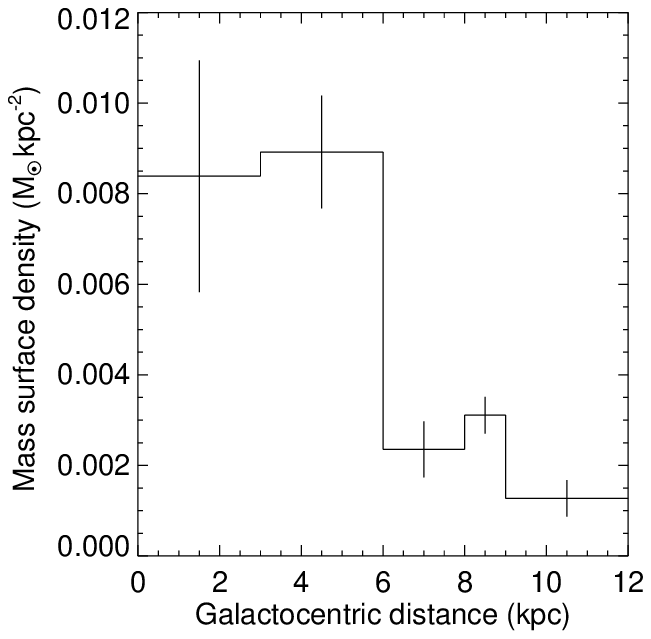}
   \hfill
   \epsfxsize=8.3cm \epsfclipon
   \epsfbox{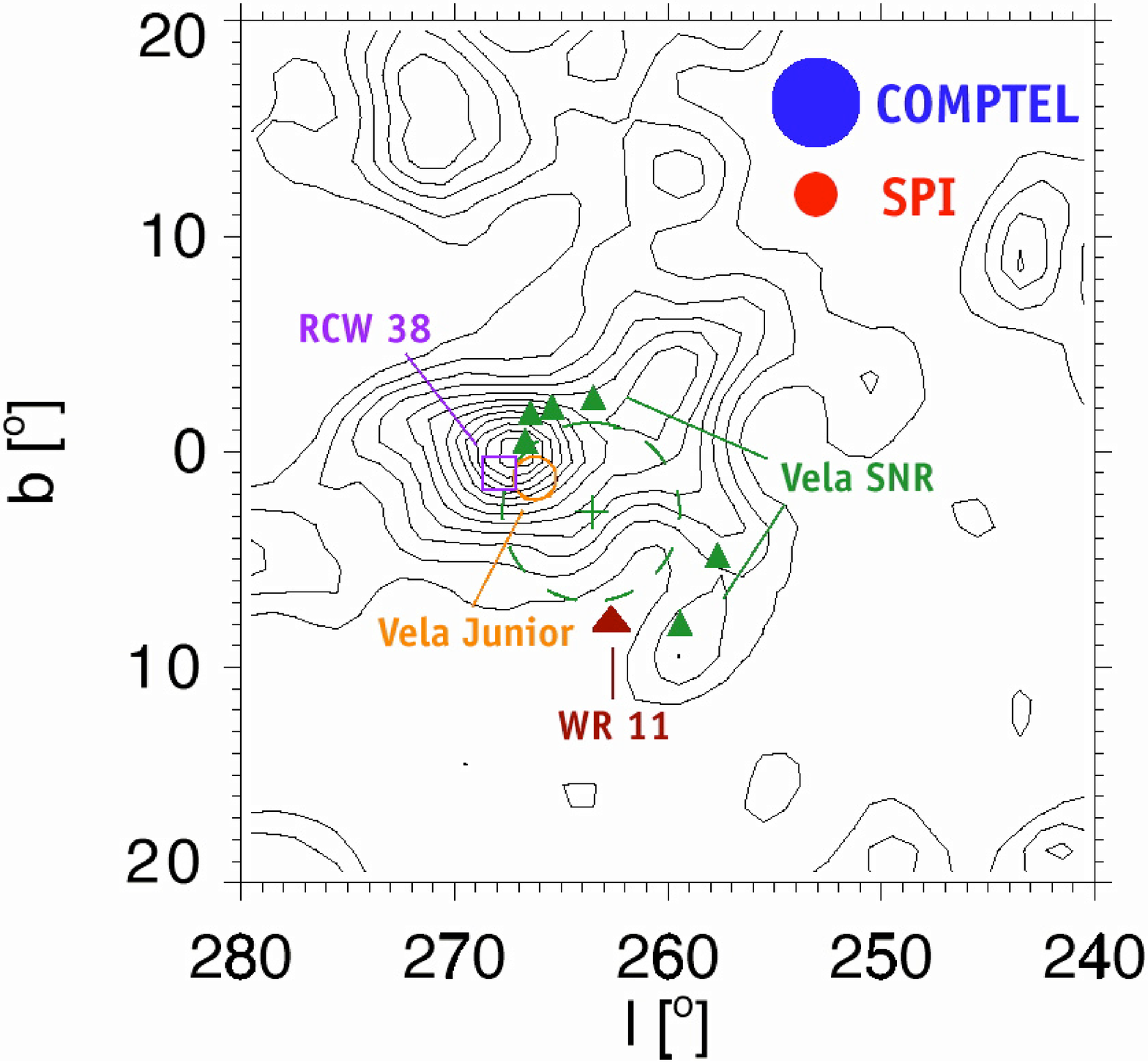}
   \caption{
   \label{fig:al26}
   Radial \al\ density profile (left) and COMPTEL 1.809 \MeV\ image of 
   the Vela region (right).
   The filled circles compare the angular resolution of the COMPTEL 
   and the SPI telescopes.
   }
\end{figure*}

Having established the correlation between 1.809 \MeV\ emission and 
massive star populations, \al\ becomes an excellent tracer of recent 
galactic star formation.
By refining the knowledge about the 1.809 \MeV\ emission distribution,
SPI will provide a unique view on the star formation activity in our Galaxy.
To illustrate this potential, the radial \nuc{26}{Al} mass density 
distribution as derived by COMPTEL is shown in the left panel of
Fig.~\ref{fig:al26}.
The bulk of galactic star formation occurs at distances of 
less than 6 kpc from the galactic centre.
Star formation is also present within the central 3 kpc of the Galaxy, 
although at a poorly determined rate.
There are indications for enhanced star formation between $3-6$ kpc, 
coinciding with the molecular ring structure.
Enhanced star formation is also seen in the solar neighbourhood 
($8-9$ kpc) which probably corresponds to activity in the local spiral 
arm.

The radial \al\ profile is probably not directly proportional to the 
radial star formation profile since \al\ nucleosynthesis may depend on 
metallicity \cite{meynet94}.
It will be important to determine this metallicity dependence in order 
to extract the true star formation profile from gamma-ray line data.
Valuable information about the metallicity dependence will come from a 
precise determination of the 1.809 \MeV\ longitude profile by SPI, and 
the comparison of this profile to other tracers of star formation 
activity, such as galactic free-free emission.
Additionally, observations of gamma-ray lines from \nuc{60}{Fe}, an 
isotope that is mainly believed to be produced during supernova 
explosions, may help to distinguish between hydrostatically and explosively 
produced \al, and therefore may allow disentangling the metallicity 
dependencies for the different candidate sources.

A precise determination of the 1.809 \MeV\ latitude profile by SPI
will provide important information about the dynamics and the mixing 
of \al\ ejecta within the interstellar medium.
High velocity \al\ has been suggested by measurements of a broadened 
1.809 \MeV\ line by the {\em GRIS} spectrometer \cite{naya96}, although 
this observation is at some point at odds with the earlier 
observation of a narrow line by {\em HEAO 3} \cite{mahoney84}.
In any case, the propagation of \al\ away from its origin should lead 
to a latitude broadening with respect to the scale height of the 
source population, and the observation of this broadening may allow 
the study of galactic outflows and the mass transfer between disk and 
halo of the Galaxy.
Actually, COMPTEL 1.809 \MeV\ observations restrict the scale height 
of the galactic \al\ distribution to $z < 220$ pc \cite{diehl97}, which
certainly excludes a ballistic motion of \al\ at a speed of 500 km s$^{-1}$.
The excellent energy resolution of SPI will easily allow to decide 
whether the 1.809 \MeV\ line is broadened or not, and the improved
angular resolution and sensitivity with respect to COMPTEL will 
allow to determine the scale height of the galactic \al\ distribution 
much more precisely.

The expected energy resolution of SPI of $\sim2.5$ keV at 1.809 \MeV\ 
converts into a velocity resolution of $\sim400$ km s$^{-1}$, allowing 
for line centroid determinations of the order of $50$ km s$^{-1}$ for 
bright emission features.
Thus, in the case of an intrinsically narrow 1.809 \MeV\ line, line 
shifts due to galactic rotation should be measurable by 
SPI \cite{gehrels96}.
Although this objective figures certainly about the most ambitious 
goals of SPI observations, a coarse distance determination of 1.809 \MeV\ 
emission features based on the galactic rotation curve seems in 
principle possible.

\subsection{Massive star clusters}

\begin{figure*}
   \epsfxsize=8cm \epsfclipon
   \epsfbox{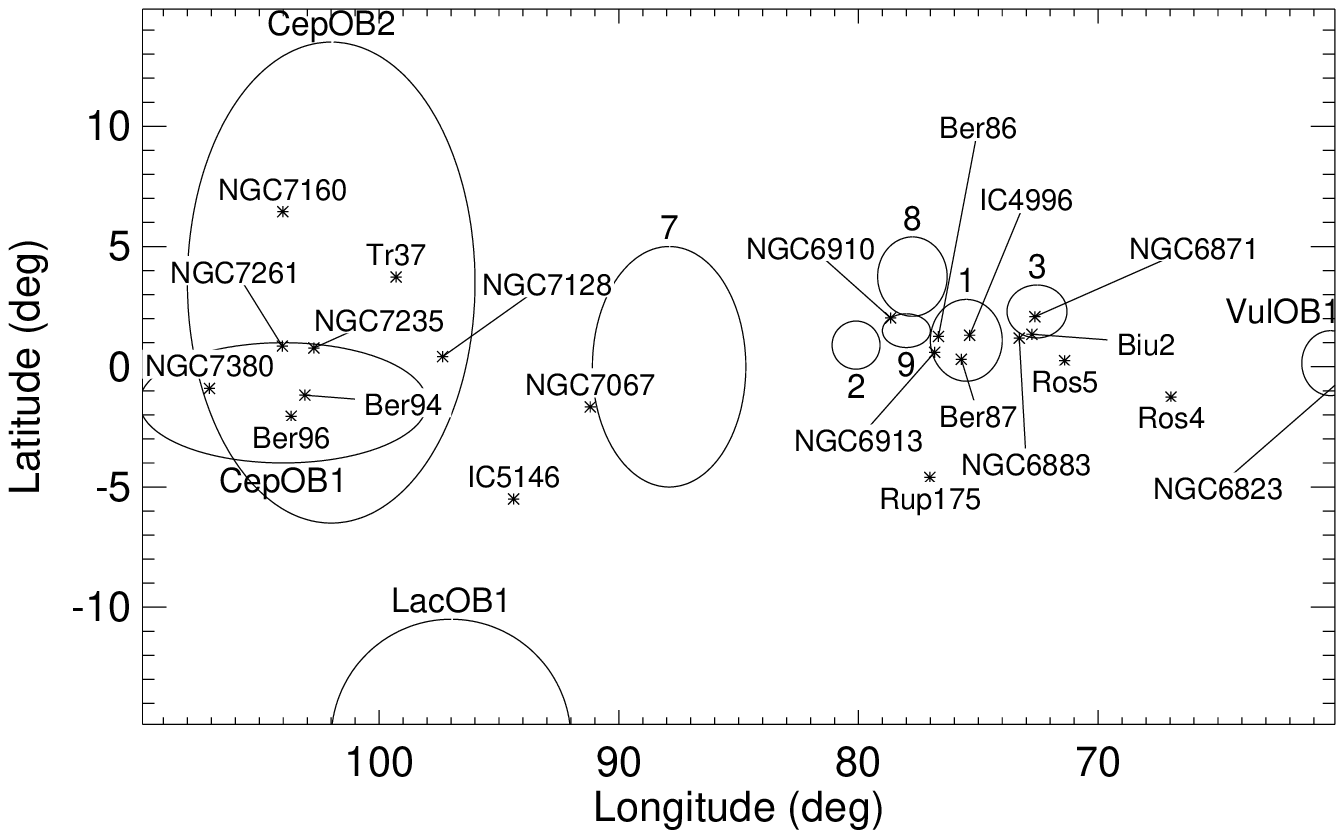}
   \hfill
   \epsfxsize=8cm \epsfclipon
   \epsfbox{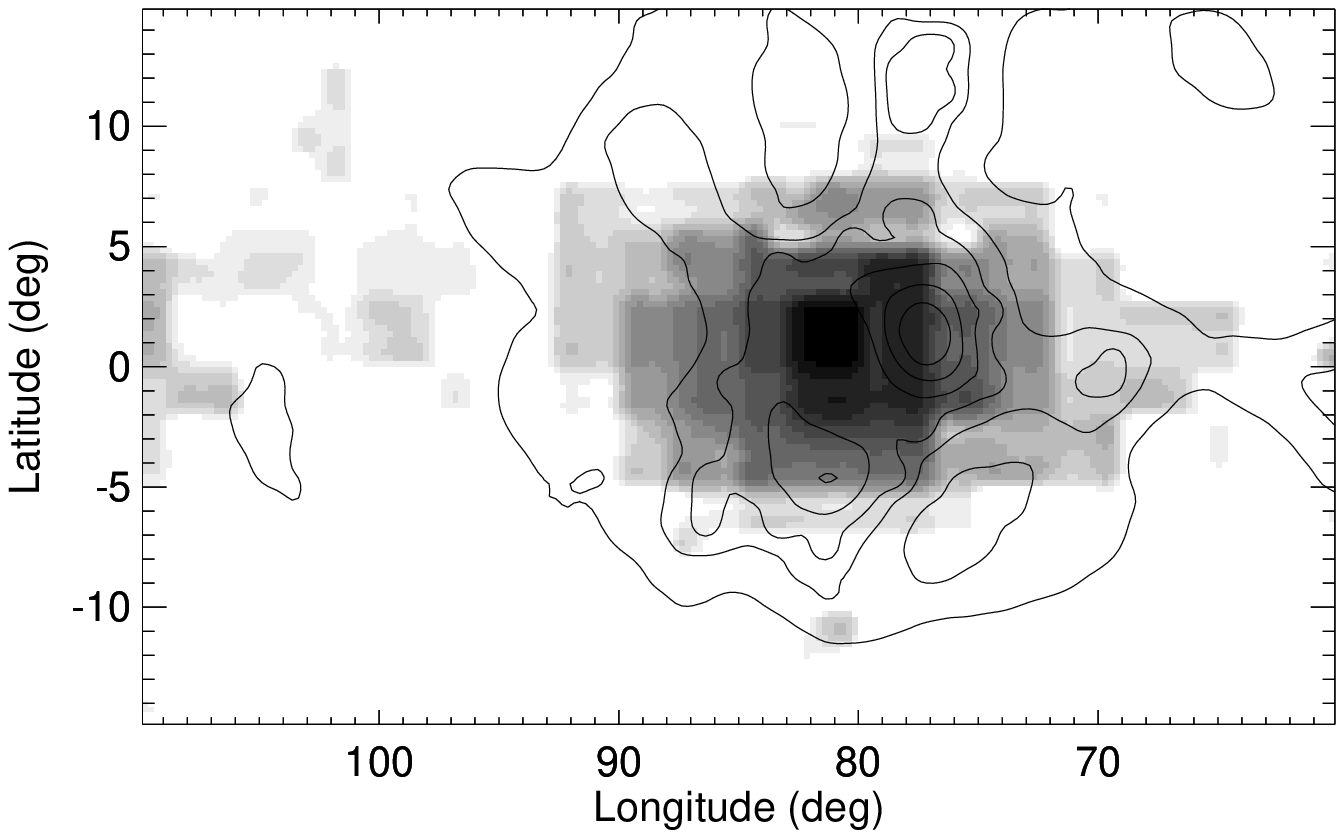}
   \caption{
   \label{fig:cygnus}
   Young open clusters (asterisks) and OB associations (circles) in the 
   Cygnus region (left) 
   and 1.809 \MeV\ COMPTEL image (contours) superimposed on a 53 GHz 
   free-free emission map of the region (right)
   (from Kn\"odlseder et al. 2002).
   }
\end{figure*}

Complementary to the study of the large-scale distribution of the 
1.809 \MeV\ emission by SPI will be observations of nearby, 
localised 1.809 \MeV\ emission regions, such as Vela, Cygnus, Carina, 
or Orion.
The aim of these observations will be the identification of emission 
counterparts at other wavelengths in order to associate the 
nucleosynthesis activity to individual objects or specific groups such 
as OB associations or young open clusters.
Already with COMPTEL, such studies have proven to provide 
important insights into the nature of \al\ sources, and 
in constraining nucleosynthetic yields for individual 
objects \cite{knoedl02}.

To illustrate the potential of SPI, a contour map of the 1.809 \MeV\ 
emission in the Vela region obtained by COMPTEL is shown in the right 
panel of Fig.~\ref{fig:al26}.
There is a wealth of potential \al\ sources in this field, but the 
limited angular resolution of COMPTEL does not allow for a clear 
identification of the dominant contributors.
Additionally, the sensitivity of COMPTEL is not sufficient to clearly 
separate diffuse from point-like emission, leading to an additional 
uncertainty in the association of emission structures with \al\ sources.
With improved sensitivity and angular resolution, SPI will help to 
overcome this problem.
Deep exposures of localised emission features will sufficiently 
constrain the 1.809 \MeV\ morphology to associate the structure with 
candidate sources in the field.
In the Vela region, which is part of the {\em INTEGRAL} core program, 
a detection of the Wolf-Rayet star WR~11 is awaited, and the 
contributions of the Vela SNR, the RX~J0852.0-4622 supernova 
remnant (Vela Junior), and OB star associations (associated to the 
radio source RCW~38) should be measurable.

In the Cygnus region, \al\ 1.809 \MeV\ gamma-ray line emission from the 
massive young globular cluster Cyg OB2 should be detectable by SPI, allowing 
for the first time the study of nucleosynthesis in an individual massive star 
association \cite{knoedl02}.
Indeed, the Cygnus region houses a variety of young stellar 
associations and clusters within an area of $\sim10\deg$ in diameter, 
separated by a few degrees (see Fig.~\ref{fig:cygnus}).
The SPI angular resolution is well matched to disentangle the 
gamma-ray line emission from these individual stellar groups, and may 
even allow to trace the propagation of the nucleosynthesis ejecta into 
the surrounding interstellar medium.
The actual COMPTEL observations (see Fig.~\ref{fig:cygnus}) indeed 
show emission maxima close to the Cyg OB2 cluster, yet the limited 
photon statistics do not allow for a detailed morphological study of 
this emission.
With its enhanced sensitivity and angular resolution, SPI should 
provide a much more detailed image of this region.
In particular, the correlation of 1.809 \MeV\ emission and free-free 
emission from the ionised interstellar medium may be studied in 
greater detail, providing a more comprehensive picture of the 
interplay between massive stars and the surrounding interstellar 
medium.

\section{Supernova nucleosynthesis}
\label{sec:supernova}

\subsection{Supernovae}

Supernovae are the most prolific nucleosynthesis sites in the Universe, 
producing a large variety of chemical elements that are ejected into 
the interstellar medium by the explosion.
Among those are radioactive isotopes that decay under gamma-ray line 
emission with lifetimes that are sufficiently long to allow escape in regions 
that are transparent to gamma-rays.
In particular, substantial amounts of \nuc{56}{Ni} and \nuc{57}{Ni} are 
produced which subsequently decay under gamma-ray line emission to 
\nuc{56,57}{Co} and finally to \nuc{56,57}{Fe}.

Observationally, type Ia events are easier to observe than the other
supernova classes because they produce an order of magnitude more radioactive 
\nuc{56}{Ni} than the other types, and because they expand rapidly enough to 
allow the gamma-rays to escape before all the fresh radioactivity has 
decayed.
From the SPI sensitivity and the observed type Ia supernova rates 
together with standard models of type Ia nucleosynthesis, one may 
estimate the maximum detectable distance for a type Ia event to about 
$15$ Mpc and the event frequency to one event each 5 years \cite{timmes97}.
Hence, during an extended mission lifetime of 5 years, SPI has 
statistically spoken the chance to detect one such event.

Nevertheless, {\em CGRO} observations have taught us that supernovae 
are intriguing objects, and even at the detection threshold, their 
observation may provide interesting implications on the progenitor 
nature or explosion mechanism.
For example, Morris {\it et al} \cite{morris95} report the detection of the 
unusually bright SN 1991T in NGC 4527 (at a distance between $13-17$ 
Mpc) by COMPTEL, implying a \nuc{56}{Ni} mass of $1.3-2.3$ \Msol.
This mass exceeds all theoretical expectations, requiring possibly a 
super-Chandrasekhar scenario to explain the observations.
Indeed, high ($\sim1$ \Msol) \nuc{56}{Ni} masses have also been inferred
from optical spectra \cite{spyromilio92,ruiz92}, suggesting a white-dwarf 
merger at the origin of the explosion \cite{fisher99}.

Another interesting example is SN 1998bu in M96, which shows the 
characteristics of a rather typical type Ia event at a distance of 
about $11$ Mpc.
Theoretical nucleosynthesis models predict that the radioactive decay 
of \nuc{56}{Co} in SN 1998bu should lead to a peak flux of 
$(1-5) \times 10^{-5}$ \funit\ in the 847 keV line \cite{gomez98}.
However, SN 1998bu was observed by OSSE for over 140 days and by COMPTEL 
for almost 90 days without any positive detection \cite{leising99,georgii00}.
The upper time-averaged 847 keV flux limit of OSSE amounts to 
$3 \times 10^{-5}$ \funit, the COMPTEL limit of $4 \times 10^{-5}$ \funit\
is comparable.
For the 1.238 \MeV\ line, COMPTEL imposes an even more stringent flux 
limit of $2 \times 10^{-5}$ \funit.
Thus, the observations start to constrain type Ia supernova models, 
excluding for example the Helium cap model for SN 1998bu 
\cite{georgii00}.
Observing SN 1998bu by SPI would probably have been a major 
breakthrough for observational gamma-ray line astrophysics.
Even if the 847 keV line would have been broadened to $50$ keV, which 
is probably rather pessimistic, SPI would have achieved a sensitivity of 
$10^{-5}$ \funit\ within a comparable exposure time ($\sim100$ days).
At this level, either SN 1998bu would have been clearly detected or 
the non-detection would have ruled out all existing thermonuclear 
supernova models.
Assuming that SN 1998bu was indeed close to detection (at a 847 keV
flux of say $3 \times 10^{-5}$ \funit), SPI would have detected the 847 keV
line at a significance of about $10\sigma$, allowing for valuable line 
profile studies.

\subsection{SN 1987A}

The explosion of SN 1987A in the Large Magellanic Cloud was a great 
opportunity for gamma-ray line astronomy.
For the first time, a supernova explosion occurred close enough to be
in reach of available gamma-ray telescopes.
The direct observation of the gamma-ray lines from \nuc{56}{Co} 
and \nuc{57}{Co} in SN 1987A was a brilliant confirmation of 
supernova theory which explains lightcurve characteristics as result of 
the radioactive decay of these isotopes.
The observed relative intensities of the gamma-ray lines from \nuc{56}{Co} and 
\nuc{57}{Co} indicated a \nuc{57}{Ni}/\nuc{56}{Ni} ratio
between $1.5 - 2$ times the solar ratio of \nuc{57}{Fe}/\nuc{56}{Fe}, 
consistent with core collapse supernova models \cite{woosley91}.

Surprisingly, the \nuc{56}{Co} lines were detected already 6 months after 
explosion, at an epoch where standard onion-shell supernova expansion models 
still predicted a substantial gamma-ray opacity for the envelope.
The gamma-ray line lightcurves presented clear evidence that \nuc{56}{Co} 
was found over a large range of optical depths, with a small 
fraction at very low depth \cite{leising90}.
Probably some fragmentation of the ejecta and acceleration of the 
emitting radioactivity are required to explain the observations.
The acceleration hypothesis is supported by various gamma-ray line 
profile measurements all indicating line widths of order 1 \% FWHM, 
corresponding to Doppler velocities of 3000 km~s$^{-1}$.

Measurements of the bolometric SN 1987A lightcurve indicate that also 
$\sim 10^{-4}$ \Msol\ of \nuc{44}{Ti} have been produced during the 
explosion, and the observation of the corresponding radioactive decay 
lines at 67.9, 78.4, and 1157 keV by SPI presents a unique tool 
to probe core collapse physics.
\nuc{44}{Ti} is synthesised in the innermost layers of the star during 
the so-called $\alpha$-rich freeze-out which is sensitive to entropy, and the 
determination of the \nuc{44}{Ti} yield would provide a direct 
measure of this quantity.
Also, the gamma-ray line profile carries valuable information about 
the velocity distribution of the matter close to the compact remnant, 
hence \nuc{44}{Ti} observations inform us about the explosion dynamics 
and possible ejecta acceleration.
Radioactive decay of \nuc{60}{Co}, accompanied by gamma-ray line emission 
at 1.173 and 1.332 \MeV, is sensitive to the neutron excess in the 
supernova, providing a unique chance for a direct measurement of this 
important parameter in SN 1987A.
The measurement of the \gray\ line fluxes will allow the determination 
of the \nuc{44}{Ti} and \nuc{60}{Co} yields in SN 1987A, providing information 
about the position of the mass-cut, the maximum temperature and density 
reached during the passage of the shock wave in the ejecta, and the neutron 
excess.

\subsection{Supernova remnants}

Until recently, the census of recent galactic supernova events was 
exclusively based on historic records of optical observations and 
amounted to 6 events during the last 1000 years.
Due to galactic absorption and observational bias, this census is by 
far not complete.
At gamma-ray energies, however, the Galaxy is transparent, and hence
gamma-ray line observations of the \nuc{44}{Ti} isotope have the 
potential to unveil yet unknown young supernova remnants.

The proof of principle was achieved by the observation 
of a 1.157 \MeV\ gamma-ray line from the 320 years old Cas A supernova 
remnant using the COMPTEL telescope \cite{iyudin94}.
Evidence for another galactic \nuc{44}{Ti} source has been found 
in the Vela region where no young supernova remnant was known before 
\cite{iyudin98}.
Triggered by this discovery, unpublished ROSAT X-ray data showing a 
spherical structure at the position of the new \nuc{44}{Ti} source 
were reconsidered and lead to the discovery of a new supernova 
remnant, now called RX J0852.0-4622 \cite{aschenbach98} (also named 
Vela Junior).
In the meanwhile, the remnant has also been discovered at radio 
wavelengths \cite{combi99}.
Although the \nuc{44}{Ti} observation is only marginal, it is the first 
time that gamma-ray line observations triggered the discovery of a new 
supernova remnant.
From the X-ray data, an age of less than 1500 yrs and a distance 
$<1$~kpc has been inferred.
Adding the \nuc{44}{Ti} observations further constrains the age and 
distance to $\sim680$ yrs and $\sim200$ pc, respectively 
\cite{aschenbach99}.
Interestingly, nitrate abundance data from an Antarctic ice core 
provide evidence for a nearby galactic supernova $680\pm20$ years ago, 
compatible with the \nuc{44}{Ti} data \cite{burgess99}.

Given the marginal detection of the 1.157 \MeV\ line from 
RX J0852.0-4622, a confirmation by SPI will be 
crucial for the further understanding of this object.
\nuc{44}{Ti} line-profile measurements will provide complementary 
information on the expansion velocity and dynamics of the innermost 
layers of the supernova ejecta.
The regular galactic plane scans and the deep exposure of the central 
radian that {\em INTEGRAL} will perform during the core program of the 
mission will lead to a substantial build-up of exposure time along the 
galactic plane, enabling the detection of further hidden young galactic 
supernova remnants through \nuc{44}{Ti} decay.
The observed supernova statistics may then set interesting constraints on 
the galactic supernova rate and the \nuc{44}{Ti} progenitors.
Indeed, actual observations already indicate that some of the galactic 
\nuc{44}{Ca} may have been produced by a rare type of supernova (e.g. 
Helium white dwarf detonations) which produces very large amounts of 
\nuc{44}{Ti} \cite{the99}.

\section{Positron annihilation}
\label{sec:positron}

The 511 keV gamma-ray line due to annihilation of positrons and 
electrons in the interstellar medium has been observed by numerous 
instruments \cite{harris98}.
At least two galactic emission components have been identified so far:
an extended bulge component and a disk component.
Indications of a third component situated above the galactic centre 
have been reported \cite{purcell97,harris98}, yet still needs confirmation 
by more sensitive instruments.

The galactic disk component may be explained by radioactive positron 
emitters, such as \nuc{26}{Al}, \nuc{44}{Sc}, \nuc{56}{Co}, and 
\nuc{22}{Na}.
The origin of the galactic bulge component is much less clear.
An apparent 511 keV flux variation from the galactic centre has led to 
the idea that a compact object might be responsible for the galactic bulge 
emission, yet the flux variation has turned out to be insignificant
\cite{mahoney94}.
Also, contemporaneous observations with the {\em SMM} satellite and latest 
observations by OSSE and {\em TGRS} show no evidence for time-variability.
This limits the flux level of any variable 511 keV point source to
less than $4 \times 10^{-4}$ \funit.
On the other hand, the observation of broadened and red-shifted 
annihilation features from 1E~1740.7-2942 \cite{bouchet91} and Nova 
Muscae \cite{goldwurm92} has been considered as evidence for positron 
production in compact objects.
However, contemporaneous observation by OSSE and SIGMA of an outburst 
of 1E~1740.7-2942 in September 1992 gave contradictory results 
\cite{jung95}, casting some doubt on the contribution of compact objects 
to the galactic positron budget.

The origin of the galactic bulge positrons will be one of the 
key-questions addressed by SPI.
Narrow-line transient features with fluxes of $4 \times 10^{-4}$ \funit\ 
should be detectable by SPI within less than one hour.
If the feature is broadened by 300 keV (as observed for example in 
1E~1740.7-2942) the required observation time increases to
$\sim12$ hours.
Hence, the weekly galactic plane scan together with the central radian 
deep exposure performed during the {\em INTEGRAL} core program will 
provide a unique survey of the galactic bulge, capable of detecting even 
faint transient 511 keV emission events.

SPI will also provide a detailed map of 511 keV emission from the Galaxy.
Using this map, the morphology of the galactic bulge can be studied in 
detail, and the question on the contribution of point sources to the 
galactic bulge emission can be addressed.
In particular, the ratio between bulge and disk emission, which is only 
poorly constrained by existing data, will be measured more precisely, 
allowing for more stringent conclusions about the positron sources of both 
components.
The 511 keV map will also answer the question about the reality of the
positive latitude enhancement, which may provide interesting 
clues on the activity near to the galactic nucleus (see von Ballmoos, 
these proceedings).

The 511 keV line shape carries valuable information about the annihilation 
environment which will be explored by SPI.
The dominant annihilation mechanism sensitively depends on the temperature, 
the density, and the ionisation fraction of the medium, and the measurement 
of the 511 keV line width allows the determination of the 
annihilation conditions \cite{guessoum91}.
Observations of a moderately broadened 511 keV line towards the 
galactic centre indicate that annihilation in the bulge mainly occurs
in the warm neutral or ionised interstellar medium \cite{harris98}.
By making spatially resolved line shape measurements, SPI will allow 
to extend such studies to the entire galactic plane, complementing 
our view of galactic annihilation processes.

With its good continuum sensitivity, SPI will also be able to detect 
the galactic positronium continuum emission below 511 keV.
The intensity of this component with respect to that of the 511 keV 
line carries complementary information about the fraction $f$ of 
annihilations via positronium formation, probing the thermodynamic and 
ionisation state of the annihilation environment \cite{guessoum91}.

\section{Conclusions}

With the research topics that we detailed in the above sections, the 
list of SPI science prospects is by far not complete \cite{knoedl00}.
Further gamma-ray line sources such as novae, accreting black holes, 
or nuclear interactions of cosmic-rays with the interstellar medium 
are valuable targets of SPI which promise new insights into 
nucleosynthesis processes and particle acceleration.
Since SPI is the spectrometer on {\em INTEGRAL} we deliberately 
excluded topics related to continuum emission -- those will be 
addressed by the corresponding paper of the IBIS collaboration 
(Lebrun, these proceedings).
Nevertheless, topics like
the diffuse gamma-ray background,
the diffuse galactic gamma-ray emission,
pulsars,
active galactic nuclei, or
gamma-ray bursts,
are equally important for SPI, which has a continuum sensitivity 
comparable to IBIS, but which is optimised for large-scale emission
(IBIS provides a much better angular resolution but is less sensitive 
to extended emission features).
On the other hand, IBIS may also detect gamma-ray lines and can help 
to identify counterparts by means of the high localisation accuracy.
Thus, SPI and IBIS are complementary instruments onboard {\em INTEGRAL}, 
which, when combined, will explore the gamma-ray sky far beyond the 
established horizon.



\end{document}